\documentclass[a4paper]{article}

\usepackage{INTERSPEECH2021}

\usepackage[utf8]{inputenc}

\usepackage{url}

\title{Speech Synthesis from Text and \\ Ultrasound Tongue Image-based Articulatory Input}
\name{Tamás Gábor Csapó$^{1,2}$, László Tóth$^{3}$, Gábor Gosztolya$^{3,4}$, Alexandra Markó$^{2,5}$}
\address{
  $^1$Department of Telecommunications and Media Informatics, \\
	Budapest University of Technology and Economics, Budapest, Hungary \\
	$^2$MTA-ELTE Lendület Lingual Articulation Research Group, Budapest, Hungary \\
	$^3$Institute of Informatics, University of Szeged, Hungary\\
	$^4$MTA-SZTE Research Group on Artificial Intelligence, Szeged, Hungary \\
  $^5$Department of Applied Linguistics and Phonetics, Eötvös Loránd University, Budapest, Hungary}
\email{csapot@tmit.bme.hu, \{tothl, ggabor\}@inf.u-szeged.hu, marko.alexandra@btk.elte.hu}

\begin{document}

\maketitle
\begin{abstract}
Articulatory information has been shown to be effective in improving the performance of HMM-based and DNN-based text-to-speech synthesis. Speech synthesis research focuses traditionally on text-to-speech conversion, when the input is text or an estimated linguistic representation, and the target is synthesized speech. However, a research field that has risen in the last decade is articulation-to-speech synthesis (with a target application of a Silent Speech Interface, SSI), when the goal is to synthesize speech from some representation of the movement of the articulatory organs. In this paper, we extend traditional (vocoder-based) DNN-TTS with articulatory input, estimated from ultrasound tongue images. We compare text-only, ultrasound-only, and combined inputs. Using data from eight speakers, we show that that the combined text and articulatory input can have advantages in limited-data scenarios, namely, it may increase the naturalness of synthesized speech compared to single text input. Besides, we analyze the ultrasound tongue recordings of several speakers, and show that misalignments in the ultrasound transducer positioning can have a negative effect on the final synthesis performance.
	
\end{abstract}

\section{Introduction}

Speech synthesis has the goal of generating human-like speech from some a specific input representation. Traditionally, this research focuses on text-to-speech synthesis, when the input is text or an estimated linguistic representation. However, a research field that has risen in the last decade is articulation-to-speech synthesis (more frequently called as articulatory-to-acoustic mapping, AAM), when the goal is to synthesize speech from some representation of the movement of the articulatory organs, without having direct access to the textual contents~\cite{Denby2010,Gonzalez-Lopez2020}. With the advent of neural vocoders, DNN-based text-to-speech synthesis has reached a mature level, i.e.~if there is a large speech database (tens of hours) available, the final synthesized speech can reach the naturalness of human communication. However, such a large database is not always available, especially when other biosignals are recorded in parallel with speech. Therefore, in limited data scenarios, DNN-TTS systems with traditional vocoders can be used. In case of articulation-to-speech mapping, there is a lack of such large databases, mainly because of the limited possibilities for recording articulatory movement in parallel with speech. Most of the articulatory recording equipment becomes highly uncomfortable for the speaker after roughly an hour. For example, recording Ultrasound Tongue Image (UTI) data requires wearing a headset, while for Electromagnetic Articulatory (EMA) recordings, cables are glued onto the tongue of the speaker. Therefore, it is worth dealing with traditional (not end-to-end) DNN-TTS methods, in case we have speech and related biosignals to process. With recent methods like WORLD~\cite{Morise2016}, MagPhase~\cite{Espic2017}, or our Continuous vocoder~\cite{Al-Radhi2020}, speech analysis and generation in statistical parametric speech synthesis has reached a mature level.

\subsection{Articulatory-to-Acoustic Mapping}

Speech sounds result from a coordinated movement of articulation organs (vocal cords, tongue, lips, etc.). The relationship between articulation and the resulting speech signal has been studied recently by machine learning tools as well. One of the research fields investigating such relationship is  articulatory-to-acoustic (forward) mapping, when the input is a speech-related biosignal (e.g.~tongue or lip movement), and the target is synthesized speech. AAM can contribute to the development of 'Silent Speech Interface' systems (SSI~\cite{Denby2010,Gonzalez-Lopez2020}). The essence of SSI is recording the articulation organs while the user of the device actually does not make a sound, but yet the machine system can synthesize speech based on the movement of the organs. In the long-term, this potential application can contribute to the creation of a communication tool for speech-impaired people (e.g. those who lost voice after laryngectomy). Voice assistants are getting popular lately, but they are still not in every home. One of the reasons is privacy concerns; some people do not feel comfortable if they have to speak loud, having others around -- but an SSI equipment can be a solution for that.

For AAM, one potential biosignal is ultrasound tongue imaging~\cite{Denby2004,Hueber2010,Csapo2017c,Csapo2020c}. For the articulatory-to-acoustic conversion, typically, traditional~\cite{Csapo2017c} or neural vocoders~\cite{Csapo2020c} are used, which synthesize speech from the spectral parameters predicted by the DNNs from the articulatory input.

\begin{figure*}%[!h]
\centering
% trims (crops) from left, bottom, right and top
\includegraphics[trim=0.0cm 25.4cm 0.0cm 0.7cm, clip=true, width=0.9\textwidth]{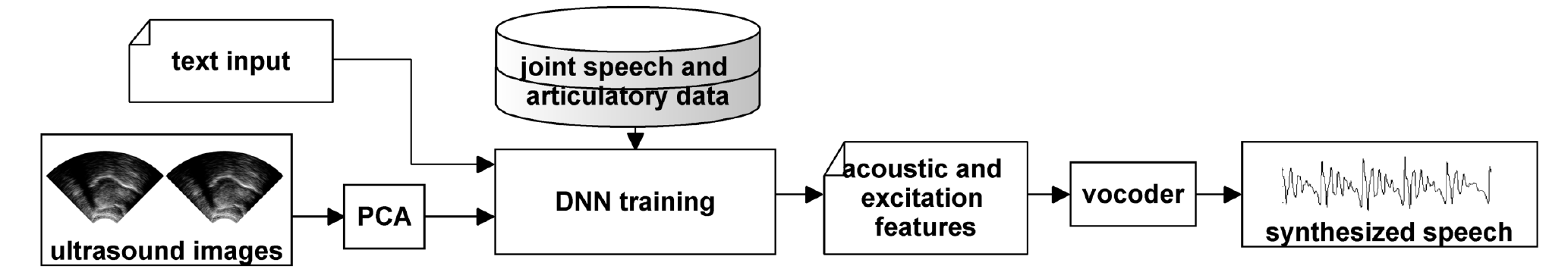}

%\vspace{-2mm}
\caption{Block diagram of the proposed approach.}
\label{fig:txtult2wav}
%\vspace{-2mm}
\end{figure*}

\subsection{Ultrasound tongue imaging}

Ultrasound tongue imaging (UTI) is a technique suitable for the acquisition of articulatory data. Phonetic research has employed 2D ultrasound for a number of years for investigating tongue movements during speech \cite{Stone1983}. Stone summarized the typical methodology of investigating speech production using ultrasound~\cite{Stone2005a}. Usually, when the subject is speaking, the ultrasound transducer is placed below the chin, resulting in mid-sagittal images of the tongue movement. Coronal images can also be acquired, depending on the orientation of the transducer. The typical result of 2D ultrasound recordings is a series of gray-scale images in which the tongue surface contour has a greater brightness than the surrounding tissue and air.
Compared to other articulatory acquisition methods (e.g.\ EMA, X-ray, XRMB, and vocal tract MRI), UTI has the advantage that the tongue surface is fully visible, and ultrasound can be recorded in a non-invasive way~\cite{Stone2005a,Csapo2017c,Ramanarayanan2018}. An ultrasound device is easy to handle and move, since it is small and light, and thus it is suitable for fieldworks, as well. Besides, it is a significantly less expensive piece of equipment than the above mentioned devices. Because of these advantages, in our study, we are using ultrasound as the articulatory information.

\subsection{TTS extended with articulatory data}
\label{sec:TTS-artic}

Articulatory information has been shown to be effective in improving the performance of HMM-based and DNN-based text-to-speech synthesis -- in an overview, Richmond and his colleagues summarize the use of articulatory data in speech synthesis applications~\cite{Richmond2015}. Ling et al. tested several ways of integrating EMA-based features into HMM-TTS~\cite{Ling2009}. They estimated the joint distribution of acoustic and articulatory features during training, by applying model clustering, state synchrony and cross-stream feature dependency. According to the results, the accuracy of acoustic parameter prediction and the naturalness of synthesized speech could be improved. Next, vowel creation~\cite{Ling2012} and articulatory control was added to HMM-TTS~\cite{Ling2013}: with an appropriate articulatory feature sequence, new vowels can be generated even when they do not exist in the training set, without using acoustic samples. The results have been also integrated into the MAGE framework~\cite{Astrinaki2013}. Cao et al.~proposed a solution to integrate EMA-based articulatory data to DNN-TTS~\cite{Cao2017}. The integration was done in two ways: 1) articulatory and acoustic features were both the target of the DNN, 2) an additional DNN represented the articulatory-to-acoustic mapping. Both naturalness and speaker identity was improved, compared to a baseline system without articulatory data.

%\cite{Ling2009}: text and articulation (EMA) $\rightarrow$ speech, HMM (increased naturalness)

%\cite{Ling2012}: vowel creation by articulatory control of HMM-TTS

%\cite{Ling2013}: articulatory control of HMM-TTS

%\cite{Richmond2015}: review, articulatory features integrated into SPSS.

%\cite{Cao2017}: integrating articulatory features (EMA) to DNN-TTS.

As shown above, integrating articulatory data to text-to-speech synthesis can improve the vocoding quality by providing more information about the vocal tract, but there is few research on this. Articulatory features derived from medical imaging data (e.g. ultrasound or MRI) have not been used before for additional input of HMM-TTS or DNN-TTS.

\subsection{Contributions of this paper}

In this paper, we extend traditional (vocoder-based) DNN-TTS with articulatory input, estimated from ultrasound tongue images. We show on the data of several speakers that this can have advantages in limited-data scenarios, in increasing the naturalness of synthesized speech compared to text input.

\section{Methods}

\subsection{Data}

We experimented with four English male (03mn, 04me, 05ms, 07me) and four female subjects (01fi, 02fe, 06fe, and 09fe) from the UltraSuite-TaL80 database~\cite{Ribeiro2021} (\url{https://ultrasuite.github.io/data/tal_corpus/}). In parallel with speech (digitized at 48~kHz), the tongue movement was recorded in midsagittal orientation using the ``Micro'' ultrasound system of Articulate Instruments Ltd. at 81.5~fps. Lip video was also recorded in UltraSuite-TaL80, but we did not use that information in the current study. The ultrasound data and the audio signals were synchronized using the tools provided by Articulate Instruments Ltd. Each speaker read roughly 200 sentences -- the duration of the recordings was about 15 minutes, which we partitioned into training, validation and test sets in a 85-10-5 ratio.

\subsection{Processing the ultrasound data}

In our experiments, articulatory features estimated from the raw scanline data of the ultrasound (i.e., echo-returns) were used as additional input of the text-to-acoustic prediction networks. We resized the 64$\times$842 pixel images to 64$\times$128 pixels using bicubic interpolation, and calculated PCA coefficients, similarly to EigenTongues~\cite{Hueber2007}. While calculating the PCA, we aimed at keeping the 70\% of the variance of the original images, thus having 128 coefficients. To be in synchrony with the acoustic features (frame shift of 5~ms), the ultrasound data was resampled to 200~Hz.

\subsection{DNN-TTS framework and DNN training}

Fig.~\ref{fig:txtult2wav} illustrates the proposed approach, i.e.~the combined articulatory and text input for the acoustic feature prediction using a DNN.
The experiments were conducted in the Merlin DNN-TTS framework~\cite{Wu2016} (\url{https://github.com/CSTR-Edinburgh/merlin}). Textual / phonetic parameters are first converted to a sequence of linguistic features as input (based on a decision tree), which are extended with the PCA-compressed version of the ultrasound tongue images. Next, neural networks are employed to predict acoustic and excitation features as output for synthesizing speech, at a 5~ms frame step with the WORLD vocoder (60-dimensional MGC, 5-dimensional BAP, and 1-dimensional LF0, with delta and delta-delta features). The DNN used here is a feed-forward multilayer perceptron architecture (six hidden layers, 1024 neurons in each). We applied tangent hyperbolic activation function, SGD optimizer, and a batch size of 256. The input features had min-max normalization, while output acoustic features had mean-variance normalization. We trained the networks for 25 epochs with a warm-up of 10 epochs, applying early stopping, and a learning rate of 0.002 after that with exponential decay. We only trained an acoustic model, and the durations were not modeled.

For baseline, we created two systems: one with text-only input, and another one with ultrasound-only input. The text-only input follows the standard Merlin recipe. The ultrasound-only input was achieved in a way that the decision tree which calculates the linguistic features was replaced with an empty tree. This way, all the remaining parameters of the training are the same in the three systems, and only the input of the networks is different.

\section{Experimental Results}
\label{sec:results}

To measure the validation and test error, we calculated both spectral prediction error (Mel-Cepstral Distortion, MCD), and excitation related errors (BAP, F0-RMSE, F0-correlation, and F0-VUV). As we only trained acoustic models, and the durations were not modeled, warping the acoustic features in time was not necessary for calculating the error measures. Several synthesized samples can be found at \url{http://smartlab.tmit.bme.hu/ssw11_txt-ult2wav}.

Table~\ref{tab:objective_MCD} summarizes the MCD results. For all speakers, the 'ult2wav' (articulatory-to-speech synthesis) system achieved the highest MCD errors (between 6.9--8.4~dB), indicating that these are relative different from the original natural utterances. The 'txt2wav' (text-to-speech synthesis) system can achieve significantly lower MCD errors, which are typically in the range of DNN-TTS with limited data (5.7--6.4~dB). Finally, the 'txt+ult2wav' (text-to-speech synthesis extended with articulatory input) system resulted in the lowest MCD scores (in the range of 
5.5--6.2~dB). According to this, adding the ultrasound-based articulatory information could enhance the prediction of the spectral features.

The results of the excitation features are summarized in Tables~\ref{tab:objective_BAP}, \ref{tab:objective_RMSE}, \ref{tab:objective_CORR},  and \ref{tab:objective_VUV}. In case of BAP (being an error difference calculated on the ban aperiodicities), the tendencies are similar as in the case of MCD: 'ult2wav' $>$ 'txt2wav' $>$ 'txt+ult2wav'. However, in case of the F0-related measures (RMSE, CORR, and VUV), the results are less straightforward. In terms of F0-RMSE, the additional articulatory input could not help during text-to-F0 prediction -- but the F0 errors with all three systems are in similar range, indicating that ultrasound itself contains some information, of which the F0 can be predicted. This is in accordance with our earlier ultrasound-to-F0 prediction experiments~\cite{Grosz2018,Csapo2019}. F0-CORR, on the other hand, is similar to MCD and BAP: here, adding the articulatory information was helpful, compared to text-only input. Interestingly, with some speakers (04me and 09fe), 'ult2wav' achieved higher correlations than 'txt2wav'. Finally, as can be seen in Table~\ref{tab:objective_VUV}, voicing can be estimated very poorly from ultrasound-only input, and adding the articulatory information to the text input did not help to improve the voiced/unvoiced decision.

Overall, we found that adding ultrasound-related articulatory information besides the textual input was useful for the spectral and BAP prediction, and in some of the F0 measures. However, there is strong speaker dependency in the results.

\begin{table}
\caption{MCD errors on the dev/test set.} \label{tab:objective_MCD}
\centering
\begin{tabular}{l||c|c|c}
     & \multicolumn{3}{c}{{MCD}} \\
Spkr & ult2wav & txt2wav & txt+ult2wav \\
\hline\hline
01fi & 8.005  / 8.094  & 5.720  / 5.636  & 5.639  / 5.565  \\
02fe & 7.674  / 7.585  & 5.974  / 5.625  & 5.767  / 5.564  \\
03mn & 7.328  / 7.153  & 5.703  / 5.652  & 5.523  / 5.442  \\
04me & 7.300  / 7.126  & 5.797  / 5.864  & 5.634  / 5.635  \\
05ms & 8.037  / 8.239  & 5.777  / 5.741  & 5.651  / 5.661  \\
06fe & 6.997  / 7.050  & 5.652  / 5.447  & 5.490  / 5.236  \\
07me & 8.426  / 8.396  & 5.989  / 5.943  & 5.851  / 5.928  \\
09fe & 7.818  / 8.351  & 6.351  / 6.566  & 6.230  / 6.439  \\
\hline
\end{tabular}
\end{table}

\begin{table}
\caption{BAP errors on the dev/test set.} \label{tab:objective_BAP}
\centering
\begin{tabular}{l||c|c|c}
     & \multicolumn{3}{c}{{BAP}} \\
Spkr & ult2wav & txt2wav & txt+ult2wav \\
\hline\hline
01fi & 0.433  / 0.428  & 0.291  / 0.269  & 0.290  / 0.276  \\
02fe & 0.311  / 0.311  & 0.246  / 0.247  & 0.241  / 0.254  \\
03mn & 0.426  / 0.402  & 0.319  / 0.322  & 0.317  / 0.323  \\
04me & 0.338  / 0.346  & 0.285  / 0.262  & 0.270  / 0.265  \\
05ms & 0.385  / 0.400  & 0.302  / 0.283  & 0.287  / 0.276  \\
06fe & 0.521  / 0.560  & 0.373  / 0.391  & 0.386  / 0.392  \\
07me & 0.689  / 0.764  & 0.437  / 0.450  & 0.454  / 0.464  \\
09fe & 0.458  / 0.511  & 0.350  / 0.397  & 0.343  / 0.394  \\
\hline
\end{tabular}
\end{table} 

%\vspace{8mm}

\begin{table}
\caption{F0-RMSE errors on the dev/test set.} \label{tab:objective_RMSE}
\centering
\begin{tabular}{l||c|c|c}
     & \multicolumn{3}{c}{{F0-RMSE}} \\
Spkr & ult2wav & txt2wav & txt+ult2wav \\
\hline\hline
01fi & 22.333 / 22.062 & 21.301 / 19.837 & 22.987 / 20.087 \\
02fe & 27.742 / 35.703 & 25.833 / 33.186 & 27.461 / 33.504 \\
03mn & 11.269 / 10.094 & 10.036 / 9.582  & 10.200 / 9.330  \\
04me & 17.809 / 23.491 & 21.672 / 28.472 & 15.955 / 22.793 \\
05ms & 11.786 / 11.892 & 11.569 / 13.208 & 10.855 / 10.724 \\
06fe & 51.407 / 40.897 & 40.784 / 39.614 & 42.861 / 39.871 \\
07me & 24.407 / 27.420 & 20.767 / 26.082 & 20.561 / 24.422 \\
09fe & 54.811 / 61.934 & 48.048 / 51.004 & 54.527 / 54.714 \\
\hline
\end{tabular}
\end{table} 

%\vspace{8mm}

\begin{table}
\caption{F0-CORR errors on the dev/test set.} \label{tab:objective_CORR}
\centering
\begin{tabular}{l||c|c|c}
     & \multicolumn{3}{c}{{F0-CORR}} \\
Spkr & ult2wav & txt2wav & txt+ult2wav \\
\hline\hline
01fi & 0.528  / 0.602  & 0.627  / 0.702  & 0.634  / 0.701  \\
02fe & 0.347  / 0.265  & 0.400  / 0.470  & 0.360  / 0.477  \\
03mn & 0.255  / 0.303  & 0.548  / 0.468  & 0.498  / 0.470  \\
04me & 0.715  / 0.741  & 0.523  / 0.423  & 0.782  / 0.745  \\
05ms & 0.550  / 0.590  & 0.565  / 0.560  & 0.649  / 0.734  \\
06fe & 0.425  / 0.657  & 0.672  / 0.649  & 0.631  / 0.652  \\
07me & 0.415  / 0.377  & 0.624  / 0.448  & 0.631  / 0.499  \\
09fe & 0.551  / 0.448  & 0.528  / 0.646  & 0.562  / 0.594  \\
\hline
\end{tabular}
\end{table} 

%\vspace{8mm}

\begin{table}
\caption{F0-VUV errors on the dev/test set.} \label{tab:objective_VUV}
\centering
\begin{tabular}{l||c|c|c}
     & \multicolumn{3}{c}{{F0-VUV}} \\
Spkr & ult2wav & txt2wav & txt+ult2wav \\
\hline\hline
01fi & 27.162 / 28.483 & 9.122  / 7.411  & 9.381  / 7.972  \\
02fe & 24.228 / 19.541 & 10.763 / 8.063  & 9.927  / 8.092  \\
03mn & 18.959 / 16.357 & 6.833  / 6.828  & 7.142  / 7.674  \\
04me & 21.597 / 22.342 & 11.602 / 9.717  & 11.320 / 10.239 \\
05ms & 26.693 / 30.381 & 11.560 / 12.669 & 12.202 / 12.929 \\
06fe & 24.201 / 21.477 & 12.217 / 7.514  & 13.079 / 8.352  \\
07me & 24.598 / 25.851 & 11.191 / 9.870  & 11.394 / 10.566 \\
09fe & 22.161 / 27.173 & 8.608  / 11.318 & 9.867  / 11.700 \\
\hline
\end{tabular}
\end{table}

\newpage

\section{Effect of ultrasound transducer position}

Next, we further investigate the strongly speaker-dependent results found in Section~\ref{sec:results}.
The articulatory tracking devices (like the ultrasound used in this study) are obviously highly sensitive to the speaker and the position of the device. A source of variance comes from the possible misalignment of the recording equipment. For example, for ultrasound recordings, the probe fixing headset has to be mounted onto the speaker before use, and in practice it is impossible to mount it onto exactly the same spot as before.  Therefore, such recordings are not directly comparable. Ultrasound-based SSI systems might not turn out to be robust against slight changes in probe positioning, which can cause shifts and rotations in the image used as input.

\subsection{Ultrasound transducer positioning and misalignment}

In order to fix head movement during the ultrasound recordings, various solutions have been proposed, e.g.~the HATS system aimed to provide reliable tongue motion recordings by head immobilization and positioning the transducer in a known relationship to the head~\cite{Stone1995}. The metal headset of Articulate Instruments Ltd.\ is a popular and well designed solution which was used in a number of studies (e.g.~articulatory-to-acoustic mapping \cite{Csapo2017c,Csapo2019}). Recently, a non-metallic system by \cite{Derrick2018} and UltraFit by \cite{Spreafico2018} are lightweight headsets to record ultrasound and EMA data. During the recording of UltraSuite-TaL \cite{Ribeiro2021}), the UltraFit headset was used.

Despite these substantial efforts, it is still a question whether the use of a headset itself is enough to ensure that the transducer is not moving during the recordings. Even if a transducer fixing system is used, large jaw movements during speech production (or drinking, swallowing) can cause the ultrasound transducer to move, and misalignment or full displacement might occur. Besides, the subjects, having discomfort due to the fixing system, sometimes readjust the headset. This way the recordings from the same session will not be directly comparable, which can be a serious issue during analysis of tongue contours. Although there exist methods for non-speech ultrasound transducer misalignment detection \cite{Narayanan2014d,Bolsterlee2016}, they cannot be directly used in speech production research.

In our earlier work~\cite{Csapo2020d,Csapo2020e}, we presented an initial idea for analyzing such misalignment. The method employs Mean Square Error (MSE) distance to identify the relative displacement between the chin and the transducer. We visualized these measures as a function of the timestamp of the utterances. Experiments were conducted on various ultrasound tongue datasets (UltraSuite, and recordings of Hungarian children and adults). The results suggested that extreme values of MSE indicate corruptions or issues during the data recordings, which can either be caused by transducer misalignment, lack of gel, or missing contact between the skin and the transducer. 

\subsection{Measuring ultrasound transducer misalignment}

The speaker-by-speaker differences of the ultrasound-to-speech conversion of the current study might also be explained with the issues of the ultrasound tongue image representation. In order to quantify the amount of misalignment, we used the MSE calculation method from our earlier study~\cite{Csapo2020d,Csapo2020e}. We compared all utterances of the eight speakers from UltraSuite-TaL with each other in the order in which they were recorded . First, for a given speaker and given session, we go through all of the ultrasound recordings (utterances), and calculate the pixel by pixel mean image (across time) of each utterance (see Fig.~1 in~\cite{Csapo2020e}). Next, we compare these mean images: we measure the Mean Square Error (MSE) between the UTI pixels ([0-255] grayscale values). MSE is an error measure, therefore the lower numbers indicate higher similarity across images. 
For a session with $n$ consecutive utterances, all compared with each other, the result is an $n \times n$ matrix (see Fig.~2 in~\cite{Csapo2020e}). We assume that if there is misalignment in the ultrasound transducer, than the matrix of measures would show this. The full details of the method, including two more similarity measures were introduced in~\cite{Csapo2020d}.

\begin{figure}
\centering
% trims (crops) from left, bottom, right and top
\includegraphics[trim=0.50cm 0.5cm 2.8cm 0.5cm, clip=true, width=0.215\textwidth]{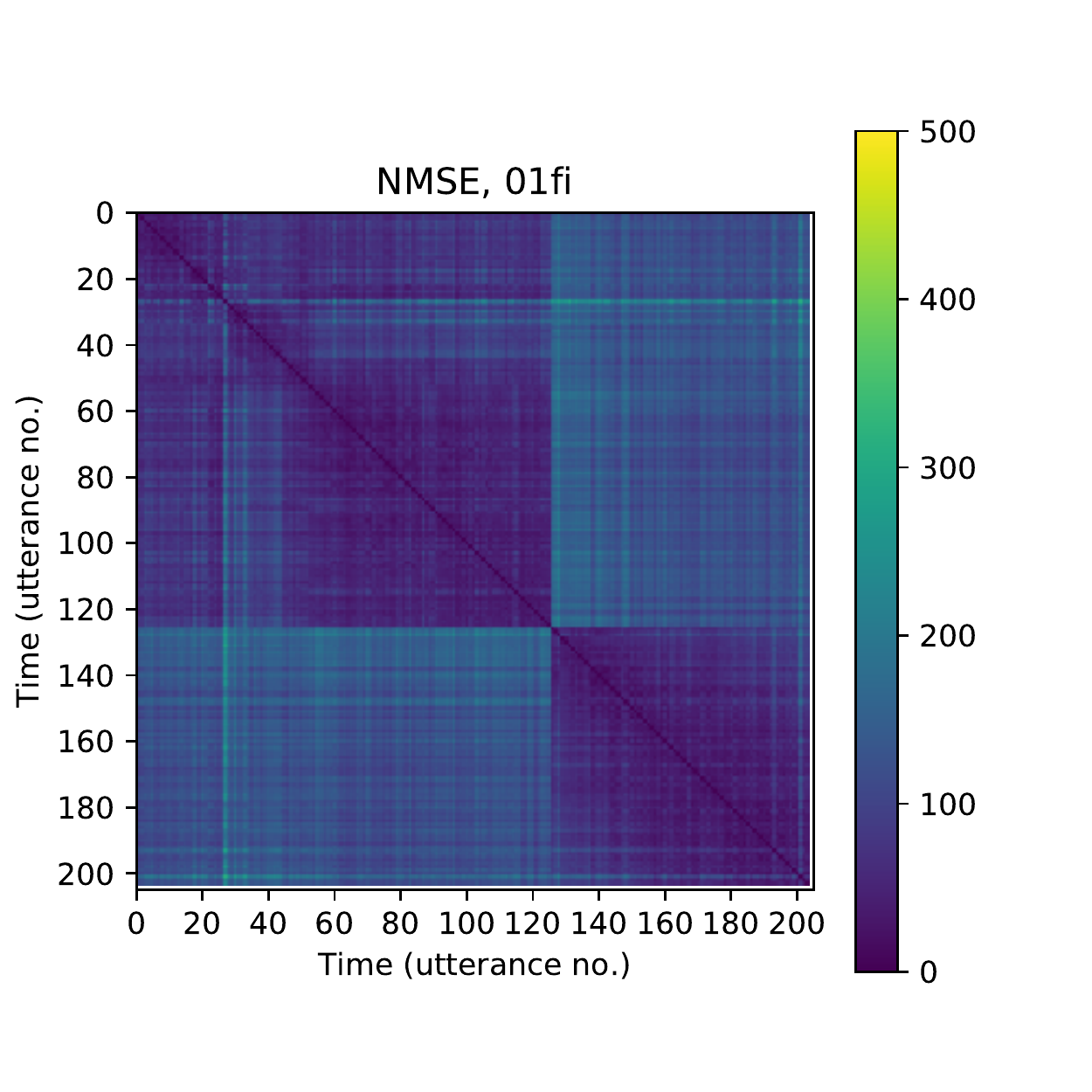}
\includegraphics[trim=1.40cm 0.5cm 0.5cm 0.5cm, clip=true, width=0.248\textwidth]{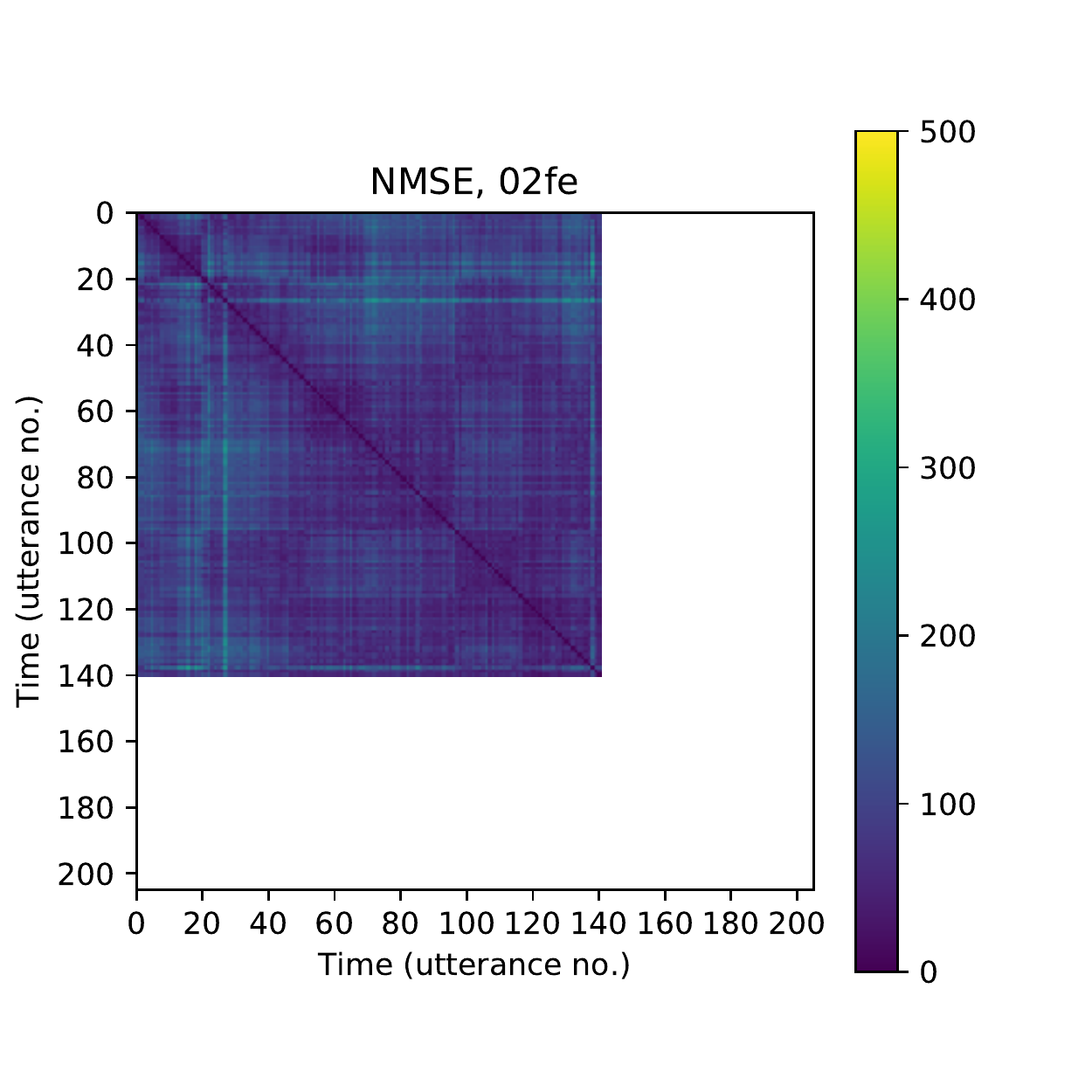}
%\vspace{-3mm}
\includegraphics[trim=0.50cm 0.5cm 2.8cm 0.5cm, clip=true, width=0.215\textwidth]{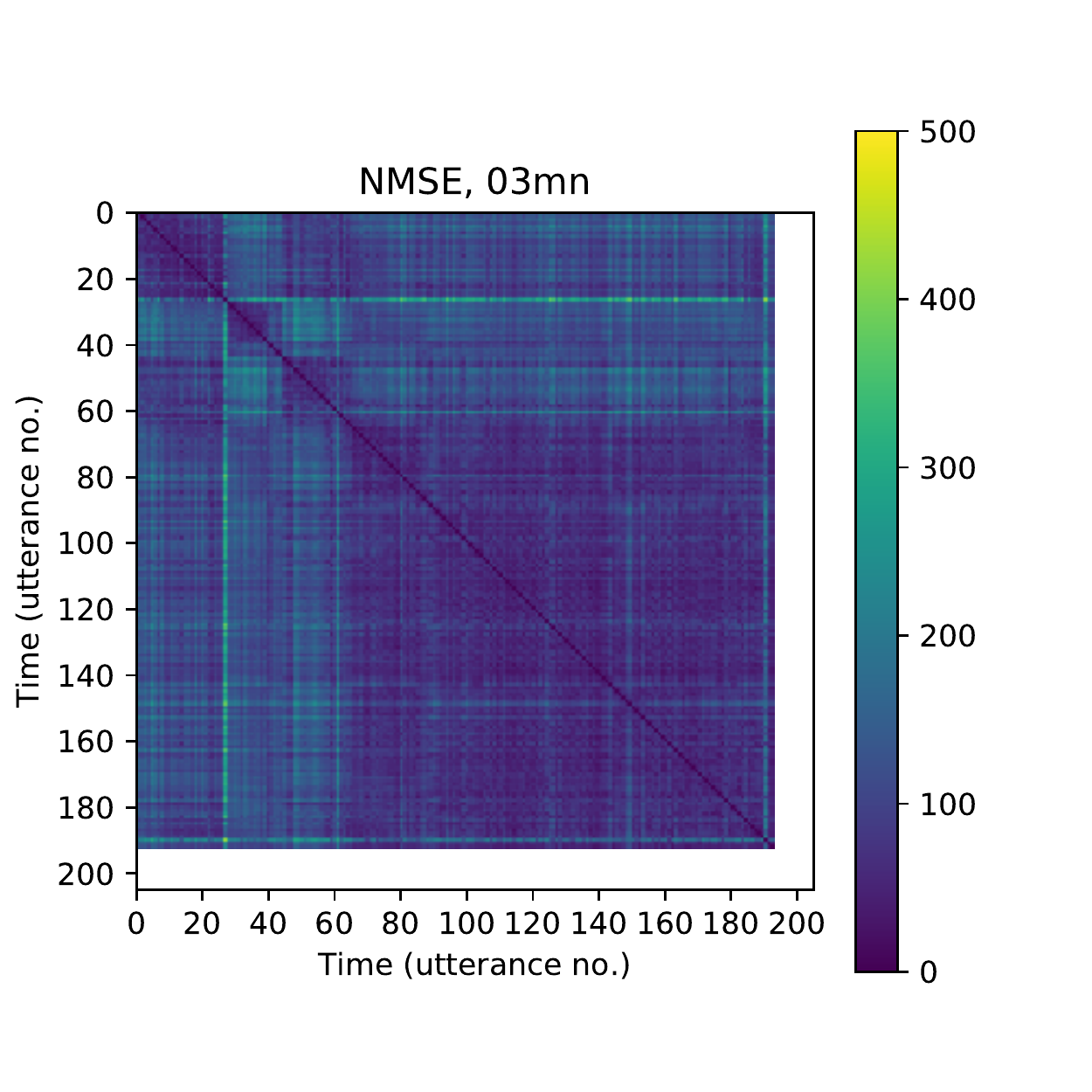}
\includegraphics[trim=1.40cm 0.5cm 0.5cm 0.5cm, clip=true, width=0.248\textwidth]{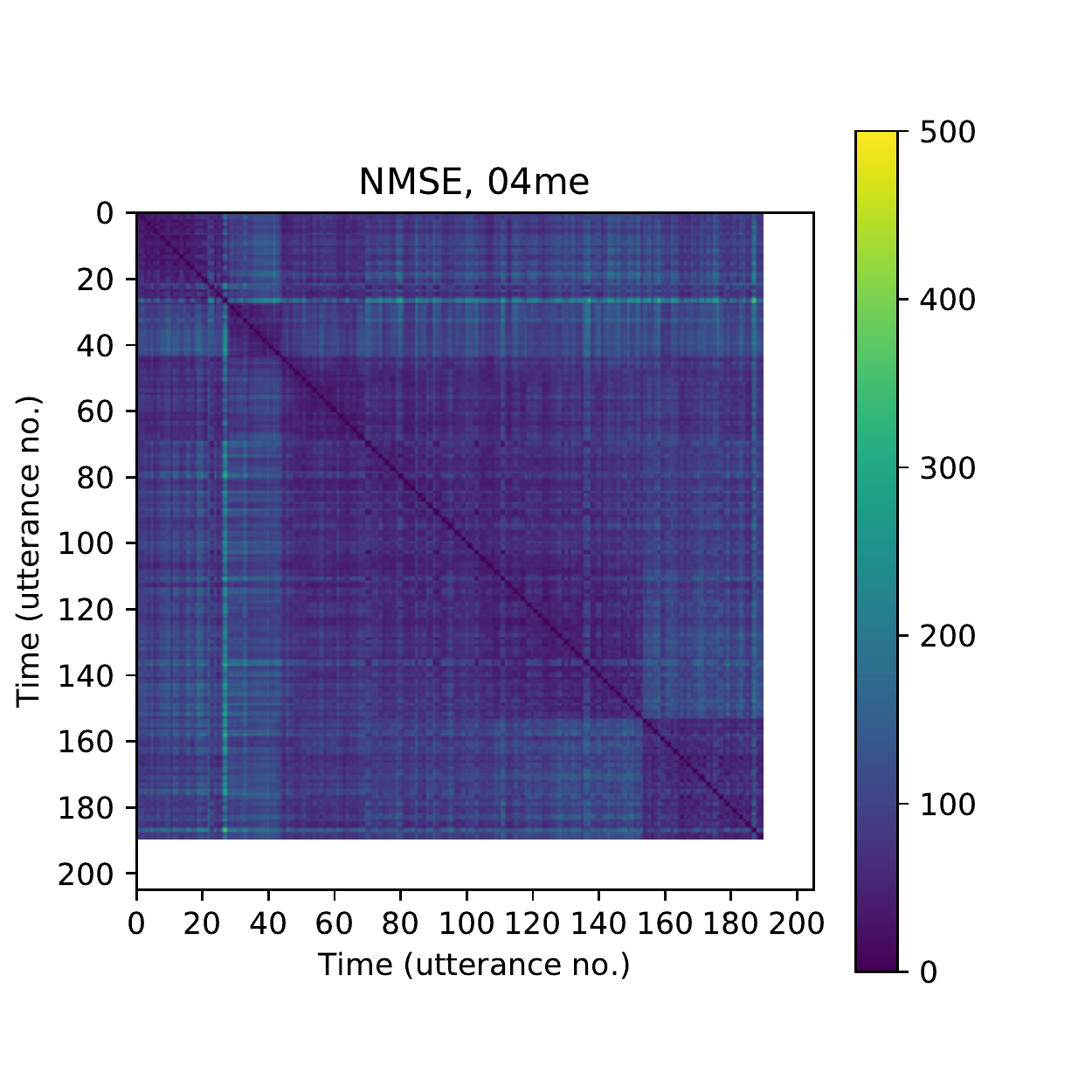}
%\vspace{-3mm}
\includegraphics[trim=0.50cm 0.5cm 2.8cm 0.5cm, clip=true, width=0.215\textwidth]{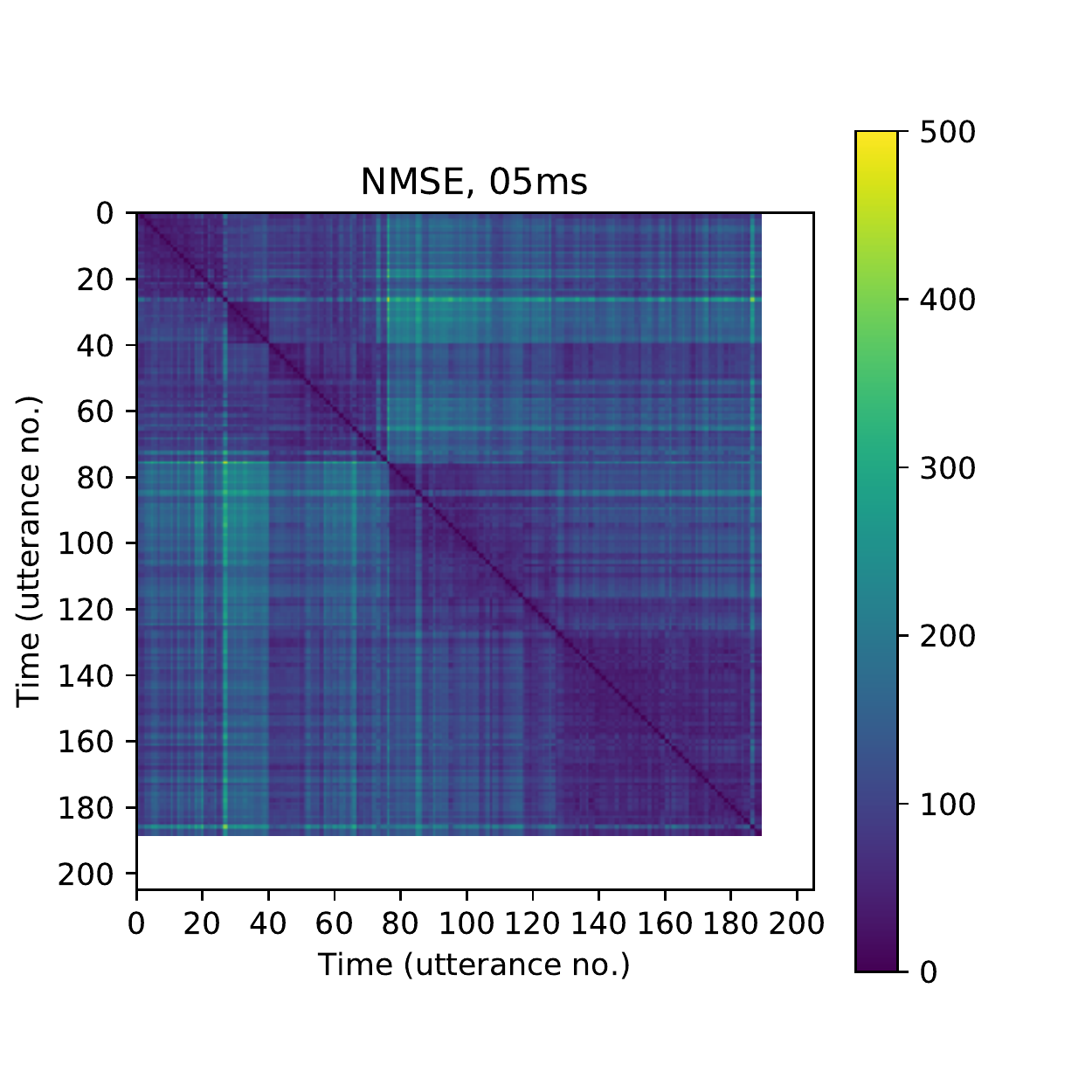}
\includegraphics[trim=1.40cm 0.5cm 0.5cm 0.5cm, clip=true, width=0.248\textwidth]{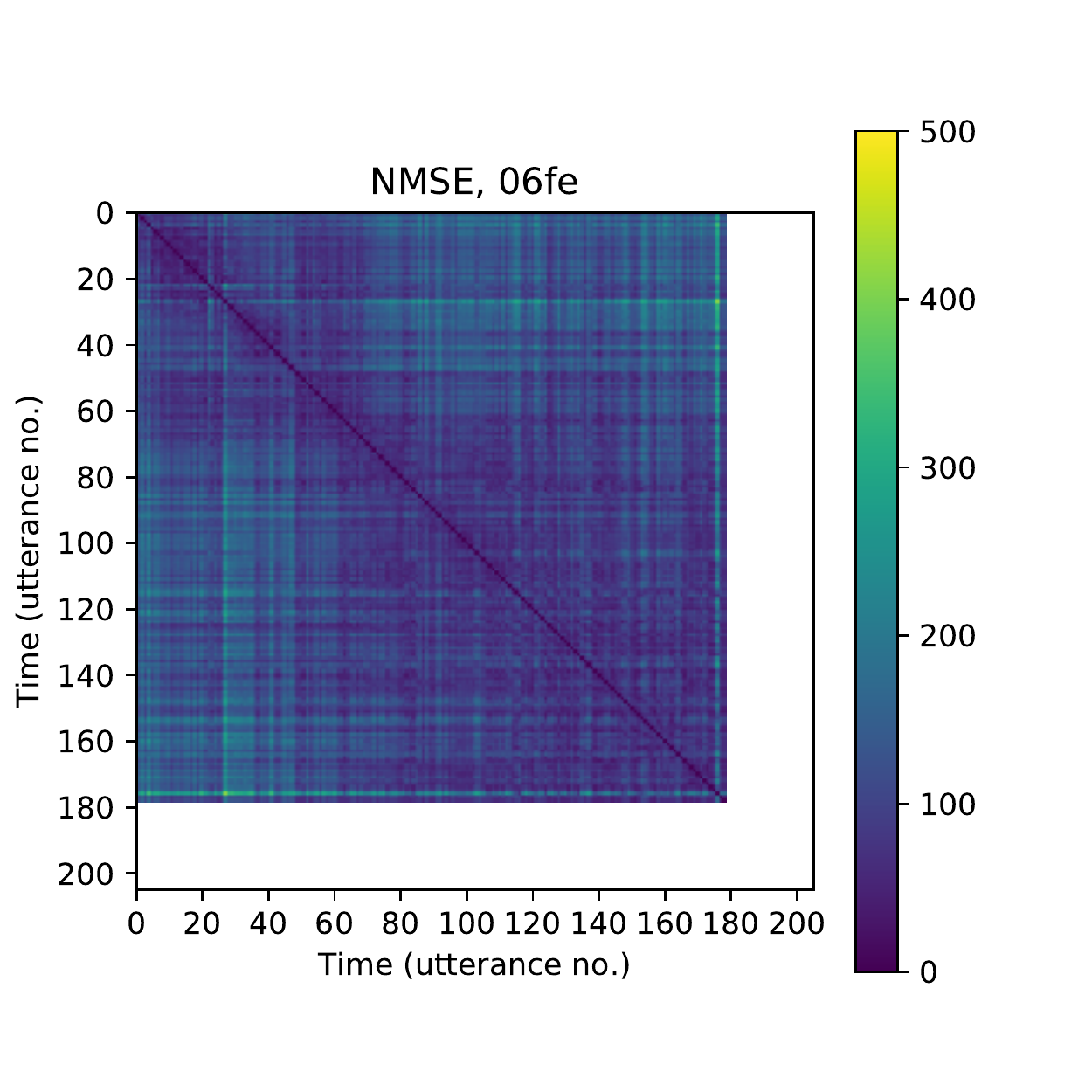}
%\vspace{-3mm}
\includegraphics[trim=0.50cm 0.5cm 2.8cm 0.5cm, clip=true, width=0.215\textwidth]{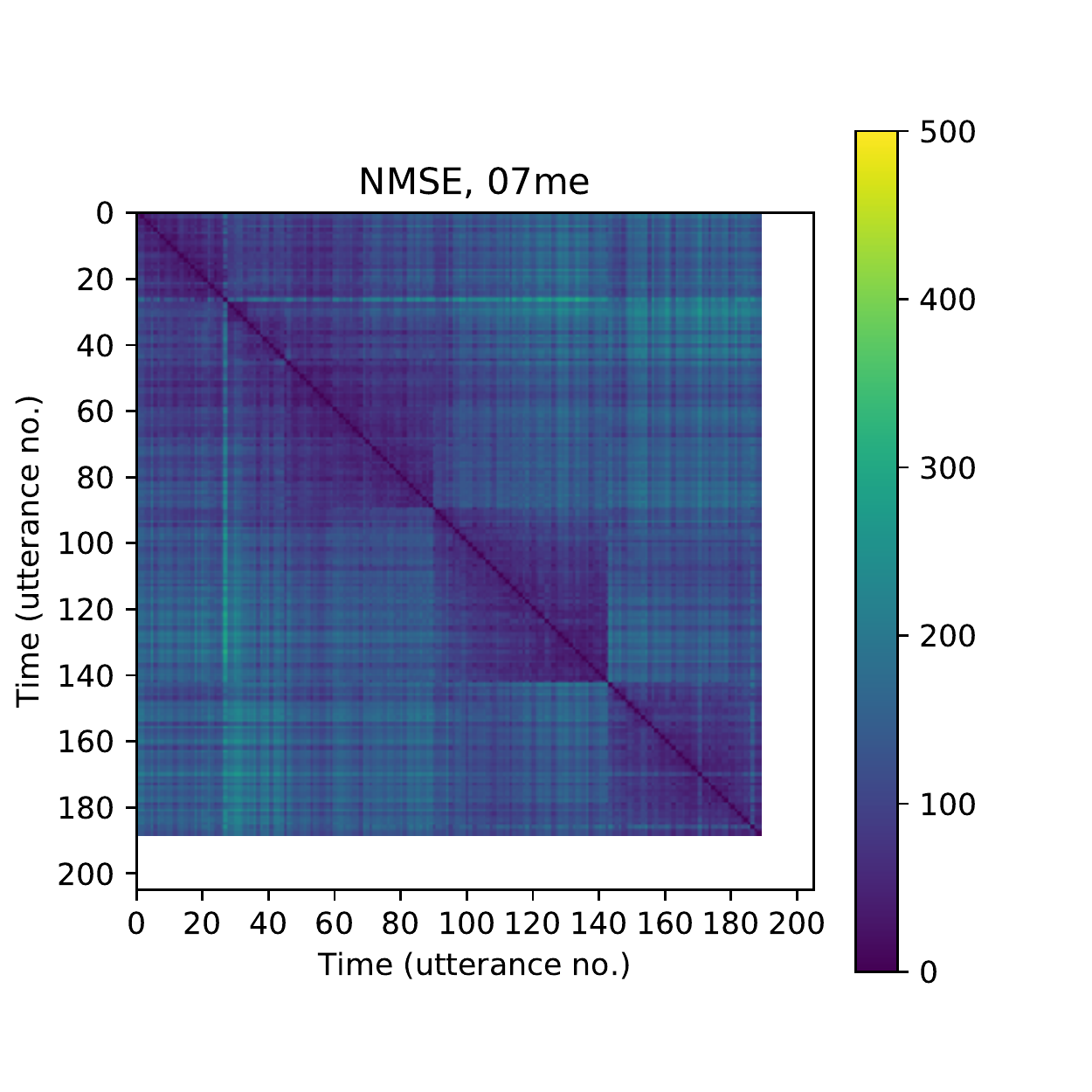}
\includegraphics[trim=1.40cm 0.5cm 0.5cm 0.5cm, clip=true, width=0.248\textwidth]{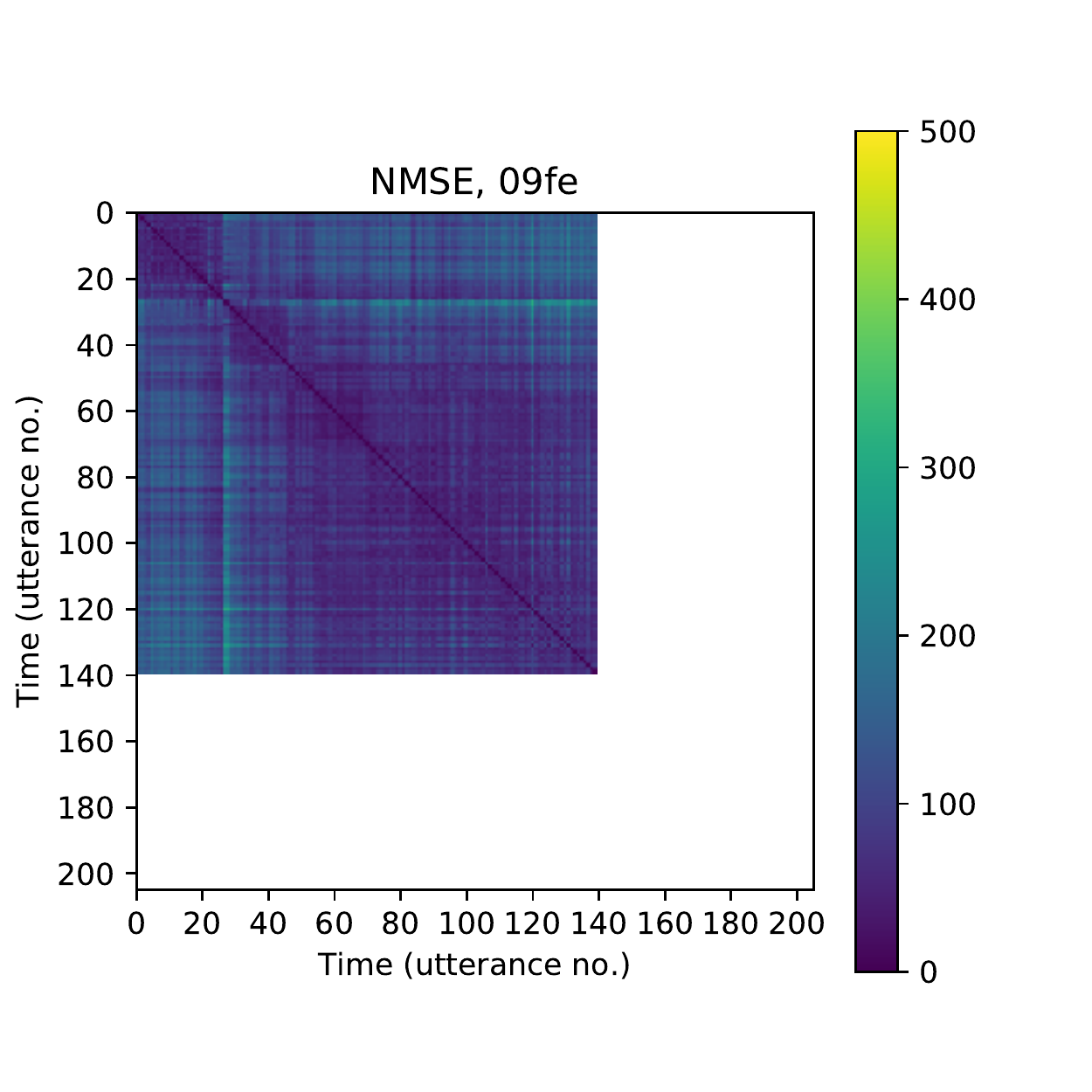}

%\vspace{-2mm}
\caption{Ultrasound transducer misalignment as a function of the utterance number within the recording session. MSE: lower values (blue colors) indicate smaller misalignment. The diagonals contain NaN values.}
\label{fig:UTI_1}
%\vspace{-2mm}
\end{figure}

The results of the ultrasound transducer misalignment MSE are shown in Fig.~\ref{fig:UTI_1}. For each speaker, the first 85\% of the data was used for training, the next 10\% for development, and the remaining 5\% for testing. On the MSE matrices of Fig.~\ref{fig:UTI_1}, the bottom left corner (or the top right corner, because the error is symmetric) indicates the differences in the positioning of the ultrasound transducer, between the training and the development/test data. If the color is yellowish, it means a higher MSE difference, i.e.~larger misalignment of the transducer. For some of the speakers, the test utterances are clearly far away (in terms of average ultrasound image) from the training utterances. For speakers 01fi, 04me, 05ms, and 07me this tendency is visible, and comparing the MSE figures (Fig.~\ref{fig:UTI_1}) with the MCD results on the development/test set (Table~\ref{tab:objective_MCD}), we can observe higher errors for them than for the remaining speakers. In case of speaker 06fe, the MSE matrix in Fig.~\ref{fig:UTI_1} is relatively homogenous, and his MCD in Table~\ref{tab:objective_MCD} is the lowest. Quantifying the exact relation between the ultrasound transducer misalignment and the acoustic / excitation errors remains future work. Also, it might be possible to auto-rotate the ultrasound images to compensate such misalignments, by comparing the actual image to an average tongue shape.

%Articulatory information has been shown to be effective in improving the performance of HMM-based and DNN-based text-to-speech synthesis -- in an overview, Richmond and his colleagues summarize the use of articulatory data in speech synthesis applications~\cite{Richmond2015}. Ling et al. tested several ways of integrating EMA-based features into HMM-TTS~\cite{Ling2009}. They estimated the joint distribution of acoustic and articulatory features during training, by applying model clustering, state synchrony and cross-stream feature dependency. According to the results, the accuracy of acoustic parameter prediction and the naturalness of synthesized speech could be improved. Next, vowel creation~\cite{Ling2012} and articulatory control was added to HMM-TTS~\cite{Ling2013}: with an appropriate articulatory feature sequence, new vowels can be generated even when they do not exist in the training set, without using acoustic samples. The results have been also integrated into the MAGE framework~\cite{Astrinaki2013}. Cao et al.~proposed a solution to integrate EMA-based articulatory data to DNN-TTS~\cite{Cao2017}. The integration was done in two ways: 1) articulatory and acoustic features were both the target of the DNN, 2) an additional DNN represented the articulatory-to-acoustic mapping. Both naturalness and speaker identity was improved, compared to a baseline system without articulatory data.

\section{Discussion and Conclusions}

In Sec.~\ref{sec:TTS-artic}, we summarized the earlier approaches that extended TTS systems with articulatory data. Most of these studies were conducted with HMMs~\cite{Ling2009,Ling2013,Astrinaki2013}, but the ideas could be applied similarly using deep neural networks, as in our experiments. All of these previous works are applying EMA as articulatory data, which is a point tracking equipment, and therefore processing that data is significantly different from the ultrasound signal that we used here. Also, the previous studies differ in the way how they include the articulatory information: it might be the input~\cite{Cao2017}, or the target of the machine learning method~\cite{Ling2009,Ling2012,Ling2013}, or also an internal representation~\cite{Cao2017}. Besides, there are many examples for DNN-based articulatory-to-acoustic mapping applying ultrasound as input, but without using the textual information~\cite{Csapo2017c,Csapo2020c,Grosz2018,Csapo2019,Kimura2019}. Although the system proposed in the current study is not suitable for direct TTS or for a Silent Speech Interface, as for the combined mapping, both text and articulatory input are required, our methods are a kind of scientific exploration, and the text-to-speech and ultrasound-to-speech results shown above might be useful for other modalities having similar properties (e.g.~rtMRI and lip images).

In this paper, we extended traditional (vocoder-based) DNN-TTS with articulatory input. The articulatory input was estimated from ultrasound tongue images, with a PCA-based compression to 128 dimensions. We have shown on the data of eight speakers from the UltraSuite-TaL dataset that this can have advantages in limited-data scenarios (e.g.~when the training data is in the range of 200 sentences for each speaker), in increasing the naturalness of synthesized speech compared to text-only or ultrasound-only input. During our experiments, we were training speaker-dependent DNNs. Creating an average voice, and adapting to a specific speaker remains future work, as it is not a trivial task. For speaker-independent training, the challenge will be to find a suitable representation of the ultrasound images, as the PCA trained on the articulatory data of one speaker is not transferable for other speakers. In the future, we plan to investigate extending DNN-TTS with other types of biosignals (e.g.~MRI or video of the lips).

The implementations are accessible at \url{https://github.com/BME-SmartLab/txt-ult2wav}.

\section{Acknowledgements}

The authors were funded by the National Research, Development and Innovation Office of Hungary (FK 124584 and PD 127915 grants). This research was supported by the project "Integrated program for training new generation of scientists in the fields of computer science", no EFOP-3.6.3-VEKOP-16-2017-00002. The project has been supported by the European Union and co-funded by the European Social Fund. We would like to thank CSTR for providing the Merlin toolkit and the UltraSuite-TaL articulatory database.

%\clearpage

\bibliographystyle{IEEEtran}

\bibliography{ref_collection_csapot_nourl}

\end{document}